\documentclass[twocolumn, amsmath,amssymb,prb]{revtex4}
\usepackage[dvips]{graphicx}

\newcommand{\ud}{\mathrm{d}}

\begin{document}

\title{The phase-locked mean impulse response \\of a turbulent channel flow}

\author{Paolo Luchini}
\affiliation{Dipartimento di Ingegneria Meccanica
Universit\`a di Salerno\\
84084 Fisciano (SA), Italy}

\author{Maurizio Quadrio}
\author{Simone Zuccher}
\affiliation{Dipartimento di Ingegneria Aerospaziale
del Politecnico di Milano \\
via La Masa, 34 - 20156 Milano, Italy}

\begin{abstract}
We describe the first DNS-based measurement of the complete mean response of a turbulent channel flow to small external disturbances. Space-time impulsive perturbations are applied at one channel wall, and the linear response describes their mean effect on the flow field as a function of spatial and temporal separations. The turbulent response is shown to differ from the response a laminar flow with the turbulent mean velocity profile as base flow.
\end{abstract}
\maketitle

Even though a turbulent flow is a nonlinear phenomenon, a linear response (either in the frequency or the time domain) can be defined for it if perturbations are small enough. We discuss in this paper the linear response of the velocity vector in the whole flow field and in time to small external perturbations applied at one wall. The vectorial nature of wall perturbations makes the response a tensorial quantity, that we denote with $H_{ij}$ to indicate the $i$-th component of the response to an impulsive wall forcing in the $j$-th direction. Wall-based forcing is particularly meaningful if the response has to be used in the context of turbulent flow control, which is the background of the present study: applying the required control at the wall is indeed the easiest configuration in a realistic flow-control setup. In view of the flow-control possibilities offered by modern MEMS technology, the linear response function can help considerably in controller design, by answering such a basic question as which effects are to be felt {\em here and now} if a wall actuator has been moved {\em there and a given time ago}.

The important role of linear processes in the self-sustaining (non-linear) turbulence cycle which takes place near the wall has been recently emphasized \cite{kim-lim-2000}. Linear control has already proven itself succesful in turbulent flows \cite{kim-2003}. Bewley and coworkers are among the most active groups in the field of linear optimal controller design for turbulent flows \cite{bewley-2001, hoegberg-henningson-2002, hogberg-bewley-henningson-2003}. They employ Kalman filters and matrix Riccati equations, which require the state equations of the system to be known. The (mean) state equations are not available, however, for a turbulent flow, and must be replaced by a linearized laminar model; in addition, the "system noise" is treated as white noise by their control design method, and the available information about the actual turbulence statistics does not enter the model. Hence these results, encouraging as they are, build on an essentially linearized laminar analysis. Replacing the parabolic Poiseuille velocity profile with the turbulent mean profile, as done by H\"ogberg, Bewley and Henningson \cite{hogberg-bewley-henningson-2003}, certainly improves the effectiveness of the controller, but the (mean) effects of turbulent mixing, which we aim at describing via the response function, still remained unaccessible to controller designers. The laminar linearized response has been recently illustrated with full detail \cite{jovanovic-bamieh-2005}. The differences between this laminar response and the mean input/output response of an actual turbulent flow will be discussed further down here.

Unfortunately, the obvious definition of an instantaneous linear response is not as useful as might be hoped in the context of turbulence: such response is bound to exhibit temporal divergence, owing to the chaotic nature of the flow. A mean response can however be given a precise meaning and measured, either experimentally or numerically. One paper which pioneered the approach is that by Hussain \& Reynolds \cite{hussain-reynolds-1970}, where the response was experimentally measured in the frequency domain at a given separation and for a few frequencies.

A few years ago, Quadrio \& Luchini \cite{quadrio-luchini-2002a} proposed a method to compute the linear impulse response function of a wall-bounded fully-developed turbulent channel flow to perturbations applied at one wall. Following that succesful proof-of-principle, we proceed here to describe and characterize the complete response function tensor, measured through a set of purposefully carried out Direct Numerical Simulations (DNS) of a turbulent channel flow. The impact that the availability of such response function will have in the field of turbulence control is the subject of ongoing work, and it has been preliminarly addressed by Luchini, Quadrio \& Bewley \cite{luchini-quadrio-bewley-2005}, who have been able to demonstrate a controller based on Wiener filtering and the present response function.

Let us consider an indefinite plane channel, bounded by two walls parallel to the homogeneous directions $x_1$ (streamwise) and $x_3$ (spanwise) and located at $x_2=0$ and $x_2=2h$. The velocity components are $u_1$, $u_2$ and $u_3$. To define an impulse response tensor, we input to the system an infinitesimal wall velocity perturbation $w_j(x_1,x_3,t) \equiv u_j(x_1,0,x_3,t)$ given by:
\begin{equation}
\label{eq:bc-impulsive}
w_j(x_1,x_3,t) = \epsilon_j \delta(x_1) \delta(x_3) \delta(t), \qquad j=1,2,3
\end{equation}
with $\delta$ denoting Dirac's delta function. The output to be measured is the mean effect of this perturbation on the velocity field throughout the channel at all subsequent times. The impulse-response tensor $H_{ij}$ so obtained relates the mean linear response of the turbulent flow to a generic input $w_j$ via the convolution:
\begin{multline}
\label{eq:linear-io}
u_i(x_1,x_2,x_3,t) = \\
\int H_{ij} (x_1-x_1',x_2,x_3-x_3',t-t') w_j(x_1,x_3,t) \ud x_1' \ud x_3' \ud t'
\end{multline}

Since turbulence fluctuations are large compared to the amplitude $\epsilon_j$ of the external perturbation, which must be small enough for the response to be linear and the relation (\ref{eq:linear-io}) to be valid, the definition of $H$ cannot be of direct use for its actual measurement. However, an ensemble average can be used conceptually to define the mean response over repeated applications of the impulsive forcing.

Our first attempt to calculating $H$ has been similar to that employed by Hussain \& Reynolds \cite{hussain-reynolds-1970}, i.e. working in the frequency domain: a DNS of a turbulent channel flow is performed, where the boundary condition is: \[
w_j(x_1,x_3,t) = \epsilon_j \sin (k_1 x_1) \sin(k_3 x_3) \sin (\omega t), \qquad j=1,2,3.
\]

Once a suitably small \endnote{Being limited to small control amplitudes is a reasonable hypothesis, as long as control is aimed at turbulent drag reduction} amplitude $\epsilon_j$ is chosen (which in general depends on the forced component), and frequency $\omega$ and wavenumbers $k_1$ and $k_3$ are given, a phase-locked average allows the deterministic effect of the perturbation to be separated from the turbulent noise with reasonable values of the signal-to-noise ratio S/N. However this numerical experiment only yields the response function in a single point of the 3d space $(k_1,k_3,\omega)$, and we soon realized that the repetition of the computation for a number of frequencies and wavenumbers large enough to yield a reasonably complete characterization of $H_{ij}$ would have been impractical.

Then we turned our attention to the direct use of (\ref{eq:bc-impulsive}) as boundary condition, with a suitably small amplitude $\epsilon_j$: from a computational viewpoint, the ensemble average can be replaced by an average over periodic repetitions well separated in time, and the complete response function is obtained at once. We again realized early that this too was going to be an unaffordable simulation: whereas impulsive forcing provides in one shot the same amount of information as many sinusoidal simulations, it does so at the expense of larger nonlinear effects. The correspondingly smaller allowed $\epsilon_j$ implies a smaller S/N ratio, and the averaging time required to bring S/N within reasonable limits becomes unaffordably long.

We eventually realized that the best of both worlds could be obtained by resorting to statistical correlation as a method for measuring the impulsive response. It is well known from signal theory that, when a white noise (i.e. a delta-correlated signal) is passed through a linear system, the correlation between input and output is proportional to the impulse response of the system. We thus adopted an indipendently generated random signal as our wall forcing, and obtained at once the whole space-time dependence of the impulse response by computing such a correlation.

In our method a DNS is performed with a zero mean white-noise signal (the output of a random-number generator) as boundary condition and the space-time correlation between this boundary condition (input) and the whole flow field (output) is accumulated. Since the applied random signal is uncorrelated to the turbulent fluctuations, the latter will be averaged out just as in phase-locking, and the deterministic response will progressively emerge while the simulation runs. Moreover, the forcing power is uniformly distributed over time and space, as opposed to what occurs in impulsive forcing, and the amplitude can be as large as with sinusoidal forcing.

\begin{figure}
\includegraphics[width=\columnwidth]{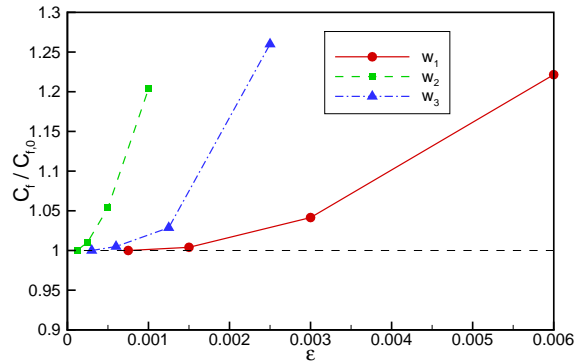}
\caption{(Color online) Effect of the white-noise power spectral densities $\epsilon_j$ on the mean friction coefficient $C_f$, for different components of the wall random forcing $w_j$. $C_{f,0}$ is the friction coefficient of the reference flow with no-slip boundary conditions.}
\label{fig:linearity}
\end{figure}

The numerical simulations are carried out with our DNS pseudo-spectral solver, whose characteristics have been described elsewhere \cite{luchini-quadrio-2006}. Of particular relevance here is the ability of the code to run in parallel with high efficiency. The Reynolds number is $Re_\tau=180$ based on the friction velocity and half the channel width. The domain size is $L_x = 4 \pi h$ and $L_z = 4.2 h$, so that 192 and 128 Fourier modes (before dealiasing) in the streamwise and spanwise directions respectively, as well as 128 point in the wall-normal direction, are required to match the commonly employed spatial resolution \cite{moser-kim-mansour-1999}. Peculiar to the present simulations is the extremely long integration time, about $10^5$ viscous time units, required to extract the deterministic response from the turbulent noise. This time interval is more than one order-of-magnitude larger than what is typically employed to obtain converged low-order statistics of the turbulent flow. The correlation is computed from products in spectral space whenever possible; its full $y$ behavior, as well as 81 time separations from $t^+=0$ to $t^+=64$ are recorded. To minimize disk space requirements, a slightly reduced set of $64$ streamwise and $84$ spanwise Fourier modes  is analysed.

A key step towards the measurement of $H_{ij}$ is the choice of the amplitude $\epsilon_j$ of the white noise applied at the wall, which must be empirically determined based on the requirement that it may yield a {\em linear} response. A preliminary estimate can be obtained by observing whether or not the forcing alters the time-mean value of the wall friction. This is shown in fig.\ref{fig:linearity}: starting from the largest $\epsilon_j$ for which (separately for each forcing component) numerical stability is preserved, we progressively halve the amplitude, and observe how the modifications of the mean friction induced by the non-homogeneous boundary condition becomes negligible: this happens only for the smallest values of $\epsilon_j$ included in the figure.

\begin{figure}
\includegraphics[width=\columnwidth]{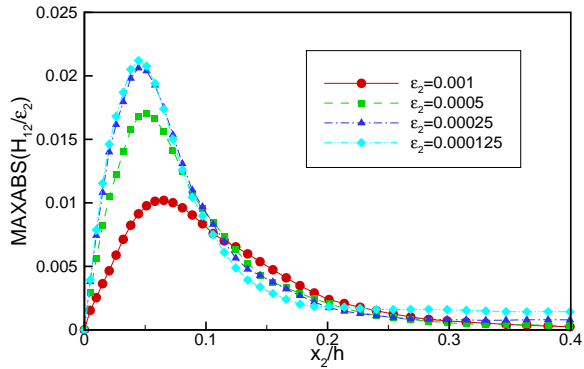}
\caption{(Color online) Variation with $x_2/h$ of the absolute-value maximum of $H_{12}/ \epsilon_2$ in wall-parallel planes. The figure refers to $H_{12}$ at $t^+=30$.}
\label{fig:miv}
\end{figure}
An actual linearity check is shown in fig.\ref{fig:miv}, where for a time delay of $t^+=30$ the maximum absolute value of $H_{12}/\epsilon_2$ in wall-parallel planes is plotted as a function of $x_2/h$. Linearity requires the curves at different $\epsilon_2$ to overlap. This is the case for the two smallest values of $\epsilon_2$, at least in the region of maximum response. At larger $x_2/h$ the curves do not collapse anymore but this is expected, since the background noise overwhelms the deterministic part of the response and the noise level is different for the various simulations (run for the same averaging time).

Having assessed linearity, and before turning to illustrate the spatio-temporal behavior of $H_{ij}$, we devote a last preliminary consideration to the response at $t=0$. The impulse response to $w_2$ includes a potential component, that can be computed analytically by solving the Laplace equation for the kinetic potential $\varphi(x_1,x_2,x_3)$ between two indefinite plane walls. The wall-normal derivative $\partial_2 \varphi$ has boundary condition $\partial_2 \varphi(x_1,0,x_3)=\delta(x_1) \delta(x_3)$. After Fourier-transforming, the problem separates into one-dimensional problems for each pair of wavenumbers $k_1$ and $k_3$. Its analytical solution reads:
\begin{equation}
\label{eq:analytical}
\widehat{\varphi}(x_2) = \frac{\mbox{Cosh} ( \kappa ( 2 - x_2) )}{\kappa \mbox{Sinh} ( 2 \kappa )}
\end{equation}
where $\kappa^2 = k_1^2 + k_3^2$.

\begin{figure}
\includegraphics[width=\columnwidth]{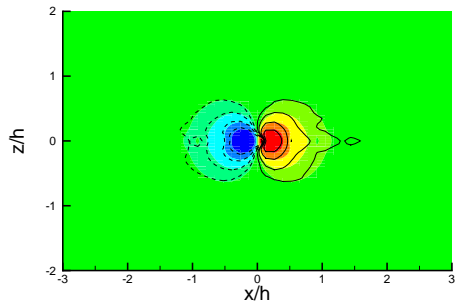}
\caption{(Color online) Streamwise derivative $\partial_1 \varphi$ at $x_2/h=0.1$ (shaded contours) of the kinetic potential, given by Eqn.(\ref{eq:analytical}), compared to $H_{12}$ at $t=0$ (line contours). Levels at $\pm$ 0.02\%, 0.01\%, 0.005\% and 0.0025\% of the maximum.}
\label{fig:potential}
\end{figure}
In fig.\ref{fig:potential} the analytical solution (\ref{eq:analytical}), represented in physical space, is compared with $H_{12}$ measured in the turbulent flow via the correlation method and shown at $t=0$: a substantially good quantitative agreement can be remarked, except for the lowest contour levels, where the residual noise becomes apparent in the turbulent response. Despite the singular nature of the potential component of the response, which manifests itself in a spike in the numerically measured correlation, this singular component is faithfully reproduced by our DNS, in which a delta-correlated boundary condition is used for $w_2$, and the correlation between this boundary condition and the whole velocity field is accumulated over time.

\begin{figure}
\includegraphics[width=\columnwidth]{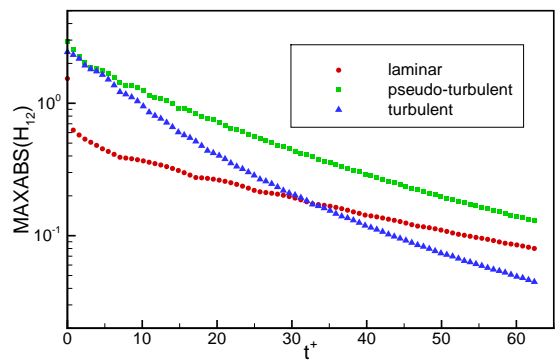}
\caption{(Color online) Decay rate of the absolute maximum value of $H_{12}$: comparison between the full turbulent response, the laminar response based on the turbulent mean profile, and the laminar response based on the Poiseuille parabolic profile.}
\label{fig:mit}
\end{figure}
We now move on to consider the spatio-temporal structure of the full tensor $H_{ij}$. We shall try to emphasize the differences between the computed turbulent response and the 'laminar' response employed until now in feedback flow-control optimization. To be precise, two kinds of linearized response have been previously considered by other authors: the solution of linearized Navier--Stokes equations about Poiseuille flow and the same solution obtained by using the actual turbulent mean velocity profile as base flow. This latter pseudo-turbulent response accounts for the mean turbulent profile but not for turbulent mixing. While the two responses with the turbulent mean profile should be identical at $t=0$ and similar at very short time delays, they can be expected to progressively diverge later owing to this difference. That this is indeed the case can be appraised from fig.\ref{fig:mit}, which reports the temporal decay of the maximum absolute value of $H_{12}$, the most frequently used component of $H_{ij}$, in the whole volume: the true turbulent response clearly presents a faster decay rate. From the same figure it can be appreciated how the laminar response is different from its companions even at $t=0$, and remains markedly different from the pseudo-laminar one for the considered time interval, thus explaining the control performance improvement observed in above cited papers upon switching from laminar to pseudo-laminar response.

\begin{figure}
\includegraphics[width=\columnwidth]{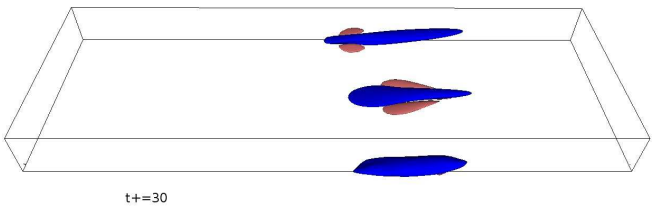}
\caption{Side-by-side comparison of $H_{12}$ for the laminar response (top), the pseudo-turbulent response (middle) and the full turbulent reponse (bottom). Isosurfaces at $\pm$ 0.7 \% of the maximum, negative values in light gray (enhanced online).}
\label{fig:comparativeH}
\end{figure}

By observing the three responses in 3d (see fig.\ref{fig:comparativeH}), other differences can be noticed. In all cases $H_{12}$ presents an elongated region of near-wall negative $u_1$, but quantitative differences are considerable. At $t^+=30$, this region turns out to be much longer and narrower for the laminar response. More importantly, both laminar responses -- and in particular the one with the mean turbulent profile -- present side regions of positive $u_1$, that are absent in the turbulent one, except for very short time delays. From the differences between the true turbulent and the pseudo-turbulent responses one is thus led to conclude that a possibility exists for further improvements in control effectiveness, if the additional information embodied in the true linear response can be exploited.

\begin{figure}
\includegraphics[width=\columnwidth]{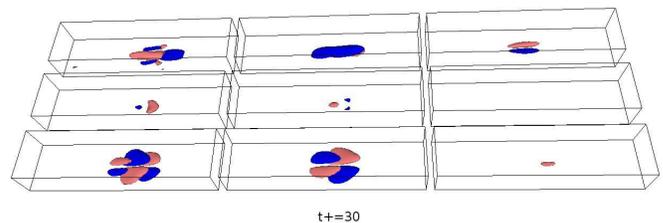}
\caption{The nine components of $H_{ij}$. Top: $H_{1j}$; middle: $H_{2j}$; bottom: $H_{3j}$. Isosurfaces of $H_{i1}$ (left) at $\pm$ 0.02 \% of the maximum; isosurfaces of $H_{i2}$ (middle) and $H_{i3}$ (right) at $\pm$ 0.2\% of the maximum. Negative values in light gray (enhanced online).}
\label{fig:H3x3}
\end{figure}

A comparative glance at all 9 components of $H_{ij}$ (fig.\ref{fig:H3x3}) reveals that they are of significantly different magnitude. The components $H_{i1}$ turn out to be relatively small, approximately 10 times smaller than the others. This parallels a similar observation  \cite{jovanovic-bamieh-2005} made for the laminar case. $H_{2j}$, i.e. the wall-normal component of the response to any forcing, decays much faster than the other components. From a qualitative viewpoint, it thus appears that the largest effect with wall-based forcing can be obtained with $u_2$ or $u_3$ actuation at the wall, and mostly $u_1$ and $u_3$ perturbations are introduced into the flow. This qualitative statement is made quantitative by the knowledge of the response tensor.

\end{document}